\documentclass[twocolumn]{aastex61}

\usepackage{color}

\newcommand\aastex{AAS\TeX}

\received{}
\revised{}
\accepted{}
\submitjournal{}

\shorttitle{\aastex\ Turbulence in the reconnection exhaust}
\shortauthors{Pucci et al.}

\begin{document}

\title{PROPERTIES OF TURBULENCE IN THE RECONNECTION EXHAUST: NUMERICAL SIMULATIONS COMPARED WITH OBSERVATIONS}

\correspondingauthor{Francesco Pucci}
\email{francesco.pucci@kuleuven.be}

\author{F. Pucci}
\affil{Center for Mathematical Plasma Astrophysics, Department Wiskunde, KU Leuven, 200B Celestijnenlaan, Leuven, 3001, Belgium}

\author{S. Servidio}
\affil{Dipartimento di Fisica, Universit\`a della Calabria, I-87036 Cosenza, Italy}

\author{L. Sorriso-Valvo}
\affil{Nanotec-CNR, U.O.S. di Cosenza, Via P. Bucci, Cubo 31C, Arcavacata di Rende, 87036, Italy}

\author{V. Olshevsky}
\affil{Center for Mathematical Plasma Astrophysics, Department Wiskunde, KU Leuven, 200B Celestijnenlaan, Leuven, 3001, Belgium}

\author{W. H. Matthaeus}
\affil{Department of physics and Astronomy, University of Delaware, 217 Sharp Lab, Newark, DE 19716, USA}

\author{F. Malara}
\affil{Dipartimento di Fisica, Universit\`a della Calabria, I-87036 Cosenza, Italy}

\author{M. V. Goldman}
\affil{University of Colorado, Boulder, Colorado 80309, USA}

\author{D. L. Newman}
\affil{University of Colorado, Boulder, Colorado 80309, USA}

\author{G. Lapenta}
\affil{Center for Mathematical Plasma Astrophysics, Department Wiskunde, KU Leuven, 200B Celestijnenlaan, Leuven, 3001, Belgium}

\begin{abstract}

%{\comm{This is just a suggestion for the abstract...:}}
The properties of the turbulence which develops in the outflows of magnetic reconnection have been 
investigated using self-consistent plasma simulations, in three dimensions.
As commonly observed in space plasmas, magnetic reconnection is characterized by the presence of turbulence. 
Here we provide a direct comparison of our simulations with reported observations of reconnection events in the magnetotail 
investigating the properties of the electromagnetic field and the energy conversion mechanisms. 
In particular, simulations show the development of a turbulent cascade consistent with spacecraft observations, 
statistics of the the dissipation  mechanisms in the turbulent outflows similar to the one observed in reconnection jets in the magnetotail, 
and that the properties of turbulence vary as a function of the distance from the reconnecting X-line.

%There results 
%These results ...are fantastic! (\comm{Here we might need to pontificate a bit ...})

\end{abstract}


\section{Introduction} \label{sec:intro}

Turbulence and magnetic reconnection are two fundamental phenomena in space plasmas. The former is responsible 
for the cascade of magnetic and ordered kinetic energy from large scale, where the energy is injected, 
to small scales, where the energy can be transformed to particle heating and acceleration. The latter consists 
in the reconfiguration of magnetic field topology with the effect of decreasing magnetic energy in favor of particle 
heating or acceleration. These two phenomena are not separate in nature, but on the contrary they 
often go ``hand-in-hand''~\citep{matthaeus2011needs}. The effects of turbulence on magnetic reconnection have been 
widely studied in magnetohydrodynamics 
(MHD)~\citep{matthaeus1986turbulent,lazarian1999reconnection,loureiro2007instability,loureiro2009turbulent,kowal2012reconnection}, 
while remaining not well-understood in the context of collisionless plasmas. 
It has also been shown that the generation of strong small-scale current sheets in the turbulent
cascade provides the conditions for the onset of reconnection, making the latter
a fundamental ingredient of the former~\citep{servidio2009magnetic,servidio2011magnetic}. 
On the other hand, the self-generation of turbulence in magnetic reconnection has been studied as 
well theoretically in two-dimensional (2D) numerical simulation of MHD~\citep{matthaeus1986turbulent,
malara1991magnetic,malara1992competition,lapenta2008self,bhattacharjee2009fast}. 
%{\comment Luca: Add something on 2-D kinetic simulations.}

In the recent years, thanks to the steady increase of the available computational resources, the 
full self-consistent description of three-dimensional (3D) reconnection has become a reality. It has been shown that in 3D new 
phenomena arise that change the picture of how and where the magnetic energy is converted to 
plasma energy~\citep{daughton2011role, lapenta2015secondary}. Many of these discoveries concern the physics of the outflows of 
reconnection:~\citet{vapirev2013formation} showed that an interchange instability develops at the interface between the plasma 
ejected from the first reconnection site and the ambient plasma;~\citet{lapenta2015secondary} that in the reconnection 
outflows a large number of secondary reconnection sites develops; and~\citet{leonardis2013identification} that 
intermittent turbulence develops in the outflows.

%\comm{Sergio: Here I would mention some observational results.} 
Observations reveal the presence of a large number of reconnection events, from large to small scales \citep{retino2007situ,greco2016complex}. Analogously, it 
is commonly observed that large scale exhausts are far from being in a laminar and regular regime, showing instead the 
clear manifestation of turbulence~\citep{osman2015multi}. More specifically, spacecraft observations of reconnection have also 
revealed the presence of turbulence within the ion diffusion region~\citep{eastwood2009observations}. 
In the inertial subrange, electric and magnetic fluctuations 
both followed a classical $k^{-5/3}$ power law; at higher frequencies, the spectral indices were near $-1$ and $-8/3$, respectively. 

In this paper, motivated by spacecraft observations, such as the studies by ~\citet{eastwood2009observations} and 
\citet{osman2015multi}, we study the properties of turbulence in the outflows of reconnection 
by means of a 3D kinetic numerical simulation.
We recover many features of turbulence that develop 
in the reconnection outflows as observed in the Earth magnetotail. 
We show that a turbulent energy spectrum develops at kinetic scale as a consequence of reconnection.
The slope of the electric and magnetic energy spectra at ion scales are found 
to be consistent with observation, along with the scale at which the two spectra depart
one from the other. We better characterize where and how the energy exchange between fields and particles 
happens in a reconnection event. Our results show that dissipation 
takes place manly in the outflows and it is intermittent. 
Moreover, we show how the properties of the turbulence varies moving away from the X-point in the outflow direction.
Our results are relevant to the physics of the magnetotail and could be useful to better understand the ongoing
Magnetospheric MultiScale (MMS) Mission observation of that region.

\section{Numerical simulation and analysis} 
\label{sec:1}
In our simulation, we consider a plasma made of protons and electrons. The initial configuration is the classical 
Harris-equilibrium:
$$ {\bf B} = B_{0x}\tanh(y/\delta){\bf e_x} + B_{0z} {\bf e_z}$$
$$ n = n_{0b} + \frac{n_{0}}{\cosh^2(y/\delta)}.$$
The coordinates are chosen as: $x$ along the sheared component of the magnetic field (Earth-Sun direction in the
Earth magnetotail), $y$ in the direction of the gradients (north-south in the magnetotail), 
and $z$ along the current and the guide field (dawn-dusk in the magnetotail).
For both species, the particle distribution function is Maxwellian with spatially homogeneous temperature.
A uniform background $n_{0b}$ is added in the form of a non-drifting Maxwellian at the same
temperature of the main Harris plasma. 
We solve the Vlasov-Maxwell equations for the two species using the semi-implicit
Particle In Cell code iPIC3D~\citep{brackbill1982implicit,markidis2010multi,lapenta2012particle}.
We consider a 3D box of shape $[40, 15, 10] \, d_p$, where $d_p$ is the proton inertial length,
which is resolved by a Cartesian grid of $[720, 270, 228]$ cells, 
each one initially populated with $125$ particles. We use a realistic mass ratio, $m_p/m_e=1836$, 
which fixes the spatial resolution to $\Delta x = d_p/18 \sim  2 d_e$, where $d_e$ is the electron inertial length. The proton 
and electron thermal speeds are $0.0063 c$ and $0.12 c$ respectively at the initial time, where $c$ is the speed of light, 
resulting in a temperature ratio of $T_p/T_e = 5$. The thickness of the initial current sheet is set to $\delta = 0.5 d_p$,
the density of the background such that $n_{0b}/n_0=0.1$, and the case of small guide field is considered: $B_{0z}/B_{0x}=0.1$.
We impose open boundary conditions in the $x$ and $y$ direction, and 
periodicity along $z$. 
The magnetic reconnection process is initialized with a 
perturbation of the $z$ component of the vector potential 
localized in the center of the domain~\citep{vapirev2013formation}.
The plasma ejected from the first reconnection sites encounters the 
ambient plasma and piles up forming the reconnection front. 
This front is unstable producing magnetic fluctuations and initiating 
the turbulent cascade. It moves towards the boundaries and 
eventually exits the simulation box.
We stop the simulation when the reconnection front 
has already started moving and is far enough  
from both the boundaries to study the turbulence that develops in front of it.
This time corresponds to $t\approx23.3\Omega_{cp}$, where 
$\Omega_{cp}$ is the proton cyclotron frequency computed using the asymptotic magnetic field,
which is the time at which we performed the bulk of our analysis.

\begin{figure}
\includegraphics[width=0.5\textwidth]{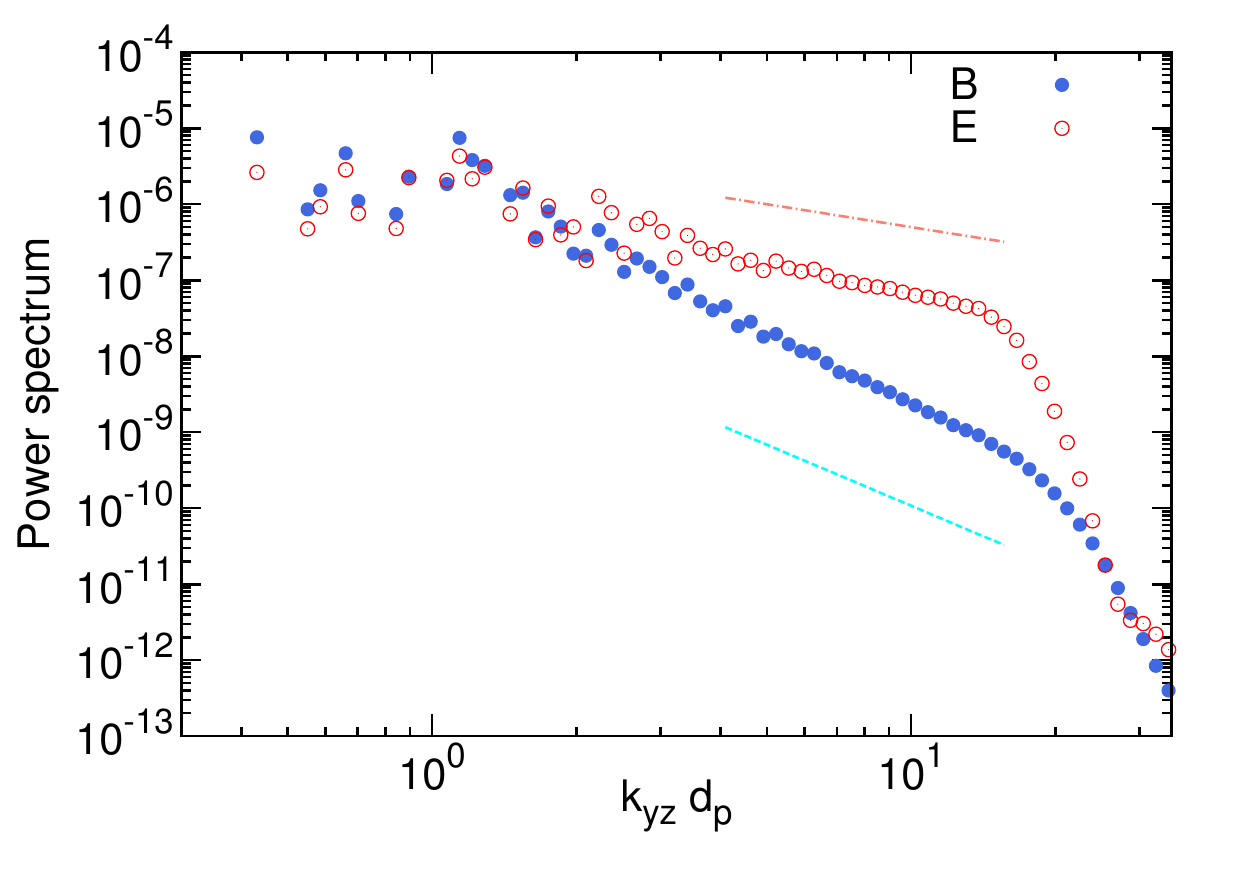}
\caption{Power spectra of magnetic (blue bullets) and electric (open red circles) fields as a function of the 
perpendicular $k$-vector (with respect to the reconnecting field direction). Spectra have been reduced along $k_x$. 
Dashed (red) line and dot-dashed (blue) lines indicate the behavior of the observations for the 
magnetic ($\propto k^{-8/3}$) and electric spectra ($\propto k^{-1}$), respectively.}
\label{fig1}
\end{figure}

\subsection{Electric and magnetic spectra}
In order to establish a first connection between plasma simulations of turbulent reconnection and 
the observations, we computed the power spectral densities of the fluctuations.
Because of the inhomogeneous background, it is important to first establish the anisotropy level and, in general, the 3D properties of turbulence.
As we are interested in the fluctuations produced by magnetic reconnection, we define the magnetic fluctuations as 
$$ 
{\bf b}({\bf x}) = {\bf B}({\bf x}) - \langle B_x({\bf x}) \rangle_{x,z} {\bf e}_x - \langle {\bf B}({\bf x}) \rangle_{x,y,z}
$$
where $\langle \bullet \rangle$ represents spatial averaging in the (2 or 3) directions indicated by the suffix. 
The above definition subtracts both mean fields and the large-scale shear: the second term in the right-hand side 
represents the signature of the background Harris sheet that is still present at the time we 
are analyzing, while the last term is the (small) guide field. The reduced auto-correlation 
function, computed in each direction, averaging over the entire volume, is defined as 
$C(r_j)=\langle {\bf b}({\bf x}+r_j\hat {{\bf r}}_j)\cdot{\bf b}({\bf x})\rangle_{x, y, z} / b^2$. Here  
$j = x, y, z$, ${\hat {\bf r}}_j$ are the unit vectors in the three directions, 
and $b^2$ the average magnetic energy of the fluctuations, i. e. $b^2=C(0)$.

In an infinite size system, for regular statistics, the autocorrelation function tends to zero for large values of the displacements, 
indicating convergence of the moments. We computed $C(r_j)$ at the peak of the nonlinear activity, measuring the correlation length of 
the fluctuations $\lambda_{C_j}$ as the displacement at which the correlation function is reduced by a factor $1/e$. Using this 
$e$-folding procedure, we measured: $\lambda_{C_x} = 4.4 d_p$, $\lambda_{C_y} = 2.8 d_p$, 
and $\lambda_{C_z} = 1.7 d_p$. This difference between correlation lengths indicates spectral (or correlation) anisotropy among the three main directions. 
The large scale (energy containing) vortexes are more elongated in the $x$ direction (along the reconnecting field). %Apart of this main axis, 
There is a secondary anisotropy due to the presence of the shear along $y$, which suggest that the coherent structures have 
the shortest extension along the periodic direction $z$. This observation is therefore in agreement with the 
anisotropic geometry of the problem.

\begin{figure}
\includegraphics[width=0.51\textwidth]{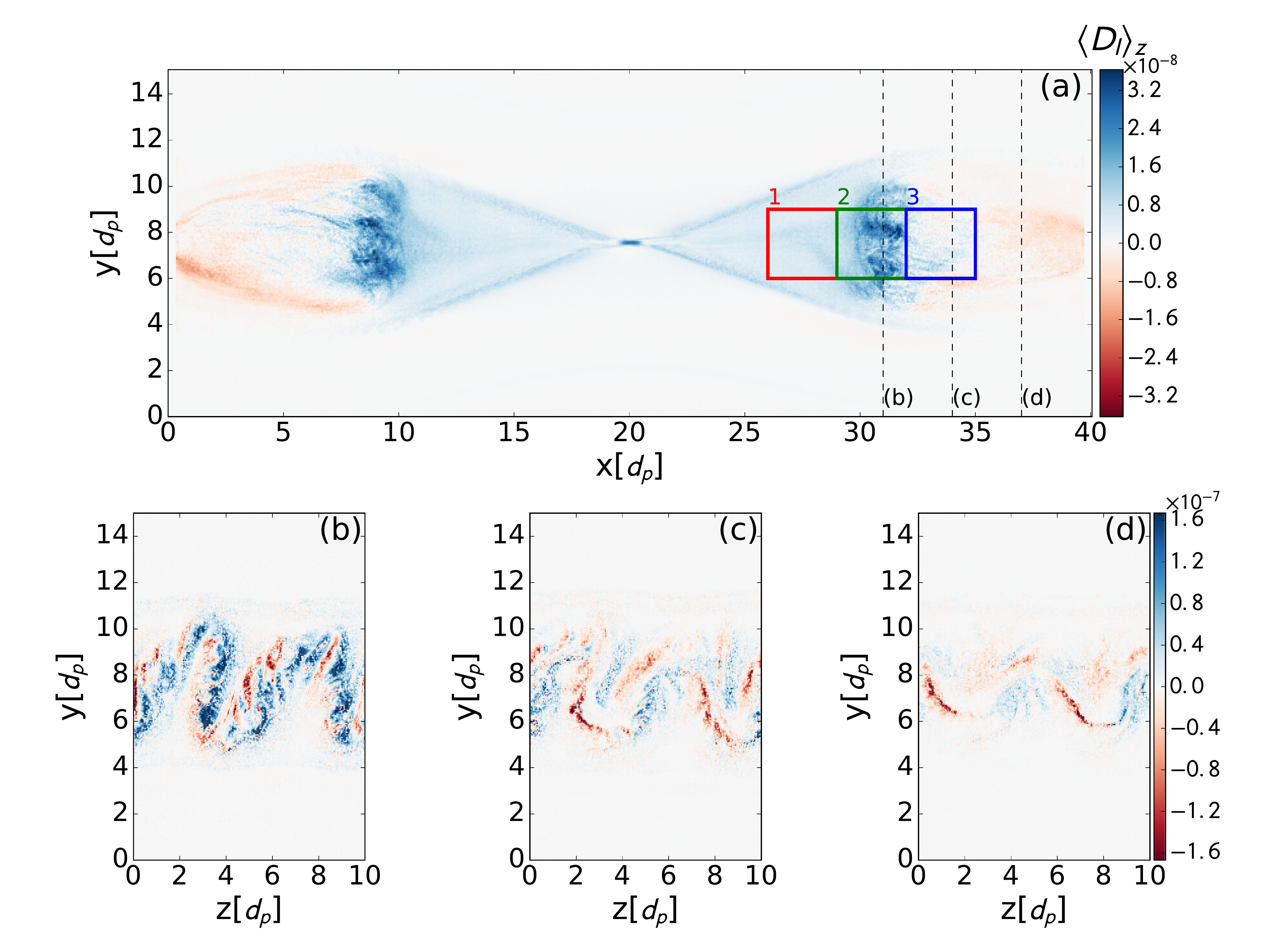}
\caption{Energy exchange $D_l = {\bf J} \cdot {\bf E}$ in the $xy$ plane averaged in the $z$ direction (a), and in the $yz$ plane at $x = 31\, d_p$ (b), 
        $x = 34\, d_p$ (c), $x = 37\, d_p$ (d). The x-line is located at $x =20 \, d_p$. The three boxes in panel (a) are the ones 
        used for the statical analysis presented in Section \ref{sec:2.2}.}
\label{fig2}
\end{figure}

In order to compute the power spectra, we used Hanning-windowing in the $x$ and $y$ directions, isolating the central exhausts. 
The window size and sharpness has been varied, verifying that the chosen parameters do not alter the 
spectrum significantly. Three dimensional energy spectra of the magnetic field
confirm what was found about the correlation lengths. It is worth noting, however, that the main anisotropy direction 
is along $x$, and that the anisotropy in the $yz$ plane is smaller, and is negligible 
at scales $r\ll\lambda_{C_z}$. This effect of isotropization is typical of small-scale turbulence.
In our case, isotropy is recovered in the $(k_y, k_z)$ plane for $k_{yz} d_p>1.5$, with $k_{yz} = \sqrt{k_y^2 + k_z^2 }$.  
Therefore it is reasonable to compute the total energy spectrum by integrating over $k_x$ and computing concentric 
isotropic shells in the $(k_y, k_z)$ plane. The results for the electric and  magnetic energy are shown in Figure~\ref{fig1}.
In agreement with space plasma observations~\citep{bale2005measurement,eastwood2009observations}, the two spectra exhibit a different power law decay 
at proton scales, with the electric spectrum proceeding at sub-proton scales with spectral index $\sim -1$, while 
the magnetic behaves more like $\sim k ^{-8/3}$.
The characteristic spatial scale at which the two spectra depart is $kd_p\sim 1$, in accordance with observations. 
In order to compare the two spectral indexes, the electric field power spectrum 
has been rescaled by a factor $5\times10^3$. ~\citet{eastwood2009observations} found a factor
of $\sim 9\times10^4$, an order of magnitude larger.
However, in our simulation the Alfv\'{e}n speed and the ion thermal
speed are $\sim 1.5$ and $\sim 3.0$ times bigger than their typical values in observation.
From a dimensional analysis of Faraday's law we get that $E/B \approx v$, where $E$, $B$ and $v$ 
are characteristic quantities. 
Hence, it is expected that the electric activity will increase if the
typical plasma velocities increase.

\begin{figure}
\includegraphics[width=0.51\textwidth]{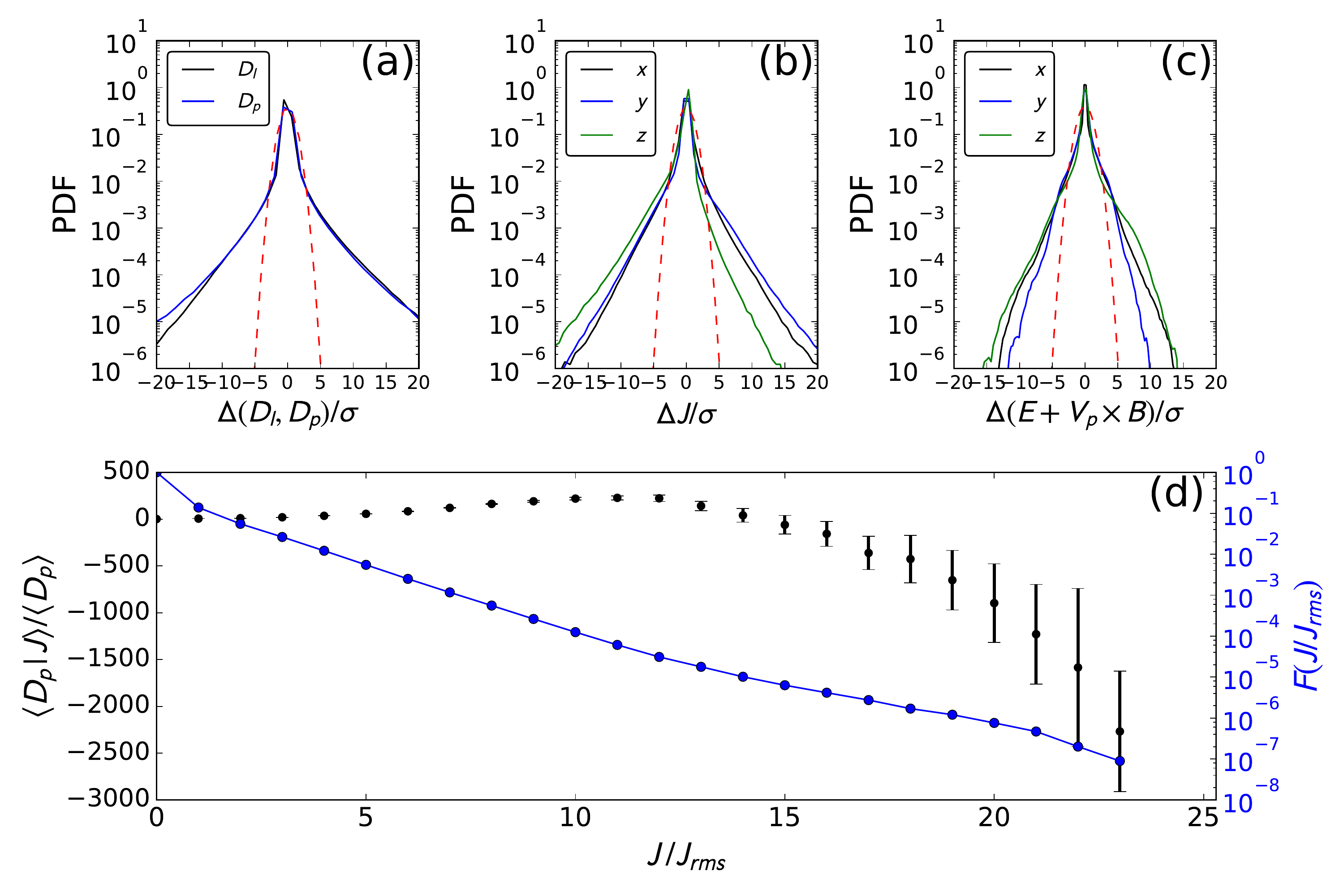}
\caption{PDFs of $D_l$ and $D_p$ (a), $\bf{J}$ (b), and $\bf{E}_p$ (c). Red dashed lines 
represent the normalized Gaussian curve. Mean $D_p$ conditioned on local current density 
thresholds and (right axis) fraction $F$ of the full box data used to compute the averages (d).}
\label{fig3}
\end{figure}

\subsection{Energy exchange between fields and particles}\label{sec:2.2}
The energy exchange between fields and particles is governed by the term $ {\bf J} \cdot {\bf E}$, 
where ${\bf J}$ is the total current, sum of protons and electrons contributions, and ${\bf E}$ 
is the electric field. When $D_l ={\bf J} \cdot {\bf E}$ is positive the energy is flowing from the 
fields to the particles, when it is negative energy is passing from particles to fields.
It is sometimes referred to as the ‘‘dissipative’’ term, even though the energy transfer from magnetic 
field to particles is not always an irreversible process, and so it does not strictly imply dissipation. Despite 
this fact, for the purpose of our paper we will keep this definition, also used 
elsewhere~\citep{zenitani2011new,olshevsky2015role,olshevsky2016magnetic}, 
and from now on we will use $D_l$ as a proxy for dissipation or more properly energy release from 
the electromagnetic field (in the laboratory frame). A 2D plot of $D_l$ 
integrated in the $z$ direction is shown in Figure~\ref{fig2}. As shown in other works, in collisionless magnetic 
reconnection $D_l$ is not concentrated only around the first reconnection site~\citep{lapenta2014electromagnetic,lapenta2015secondary}. 
In fact, it takes nonzero values in a wider region contained in the outflows (panel (a)). Moreover, 
$D_l$ is strongly inhomogeneous inside the outflows. In order to characterize this inhomogeneity we 
plotted $D_l$ in the plane facing the outflows, $yz$, at three different positions 
along $x$: $31 d_p$,  $34 d_p$, and $37 d_p$. The largest values of $D_l$ are found in the region where
the plasma ejected by reconnection encounters the ambient plasma and is decelerated, near $x = 31 d_p$ in the right outflow.
This region is characterized by an interface instability, 
which was studied in~\citet{vapirev2013formation}. $D_l$ is stronger in that position and decreases 
moving outwards from the first reconnection site in the outflow direction. Note that $D_l$ 
has in general both positive and negative values, but in the considered region its average is always positive, 
indicating a net flow of energy from fields to particles.
\begin{figure}
\includegraphics[width=0.51\textwidth]{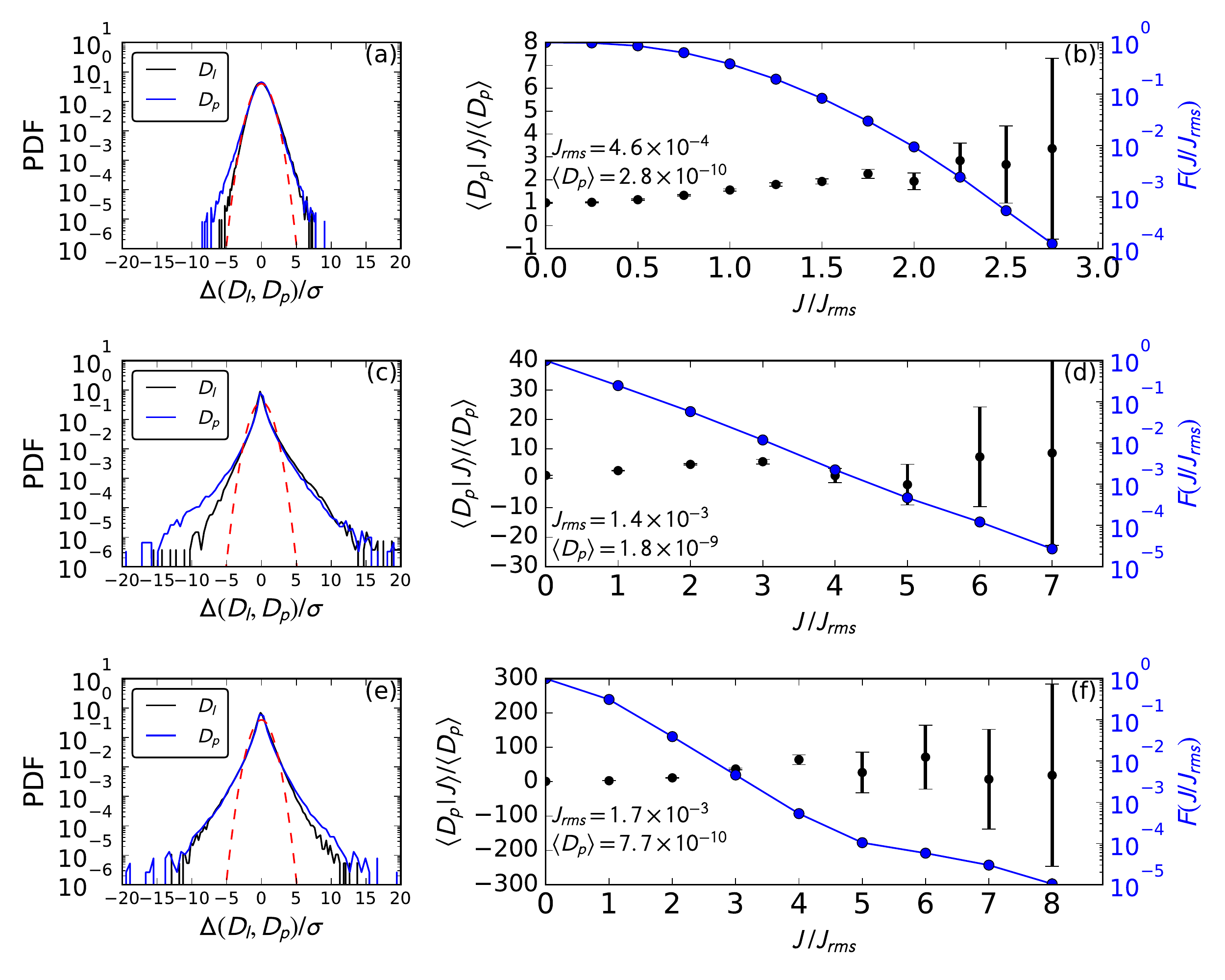}
\caption{PDFs of $D_l$ and $D_p$ in $BOX_1$ (a), $BOX_2$ (c), and $BOX_3$ (e). 
Conditioned average of $D_p$ and filling factors $F$ in $BOX_1$ (b), $BOX_2$ (d), and $BOX_3$ (f).}
\label{fig4}
\end{figure}
In order to give a better description of how energy is converted and to compare our results with observations, 
we performed a statistical analysis similar to what presented by~\citet{osman2015multi}, where a dissipation analysis 
was performed on observational data in a magnetic reconnection outflow 
in the magnetotail. It is worth noticing that in~\citet{osman2015multi} 
the statistics were made from temporal data collected by the satellites crossing the reconnection outflow, 
while in our case the whole simulation box is used as a single sample.
\begin{figure}
\includegraphics[width=0.48\textwidth]{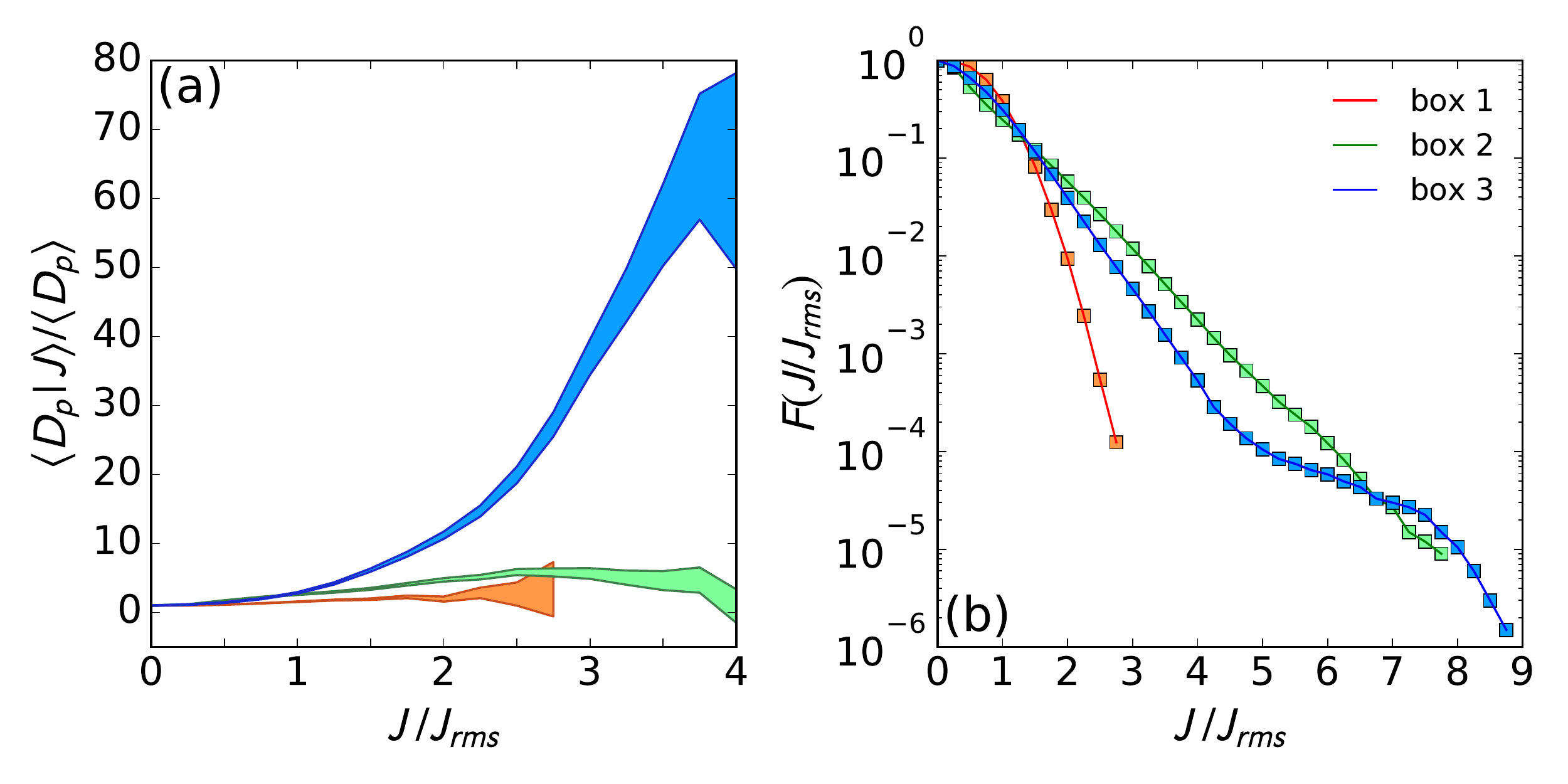}
\caption{Conditioned average of $D_p$ (a) and filling factor (b) in BOX 1 (red), BOX 2 (green), and BOX 3 (blue).}
\label{fig5}
\end{figure}
In panel (a) of Figure~\ref{fig3} the Probability Density Functions (PDFs) of $D_l$ and $D_p$ are plotted. 
$D_p$ represents the dissipation proxy in the proton reference frame and is given by
$D_p = {\bf J} \cdot {\bf E}_p$, where ${\bf E}_p = \left({\bf E} + {\bf V}_p \times {\bf B} \right)$ 
is the electric field in the proton reference frame, ${\bf V}_p$ representing the proton bulk velocity. 
The two PDFs are compared with the normalized Gaussian distribution (plotted in dashed-red line). They 
strongly depart from Gaussian distributions, presenting instead high tails up to several standard 
deviations $\sigma$, which can be interpreted as a signature of intermittency. In panel (b) and (c), we present 
separately the PDFs of the two terms which compose $D_p$. In agreement with what was found in the space measurement, 
the PDFs of ${\bf J}$ and ${\bf E}_p$ are both non-Gaussian. In panel (d), the average $D_p$ conditioned to a 
threshold current density is shown. The plot is constructed as follows: a threshold in the current density 
magnitude is considered and the average of $D_p$ is computed using all those points in the domain where the 
value of the current is bigger than the fixed threshold. This average is then normalized to the average 
of $D_p$ on all points, which gives by definition $\langle D_p | J=0 \rangle/\langle D_p \rangle = 1$.
The black points in the plots represent the result of such computation for different values of the threshold. 
The blue curve represents the filling factors, i.e. the fraction of points used for computing the average 
with respect to the total number of points in the sample.
The average of $D_p$ strongly increases when higher threshold are considered up to $J/J_{rms} = 10 $. Beyond 
this threshold, the average starts decreasing and changes sign to reach large negative values for the strongest current density.
Our results confirms that the exchange of energy is local, with larger values of $|D_p|$
localized in very small volume filling structures. We show as well that even if the average of $D_p$ is 
positive, points where the value of the current is very strong can be site of negative $D_p$. This is not
 a universal behavior, but indeed depends on the particular time considered in the simulation: we performed 
the same analysis at a different time step (not shown) finding positive values of $D_p$ for big current 
values. However, at all times the energy exchange is consistently concentrated in small regions, where 
the values of $D_p$ is much larger than its global average. 

The above statistical analysis was performed considering the whole simulation box, providing information 
about the average properties of the dissipation proxy in the reconnection events. In order to obtain a 
more detailed description, and to identify in which place the energy exchange actually occurs we 
perform the previous statistical analysis in three different regions of the simulation, by selecting 
three boxes located in the left reconnection outflows at three different distances from the X-point. 
The boxes are identified by $6<y/d_p<9$ (reconnection zone), $0<z/d_p<10$ (full domain in $z$), and with
${\mbox {BOX}}_1=\left\{26<x/d_p<29\right\}$, ${\mbox {BOX}}_2=\left\{29<x/d_p<32\right\}$, 
${\mbox {BOX}}_3=\left\{32<x/d_p<35\right\}$ (see  Figure~\ref{fig2}). 
Moving further in the direction of the outflows the PDFs become non-Gaussian. This transition happens 
between $BOX_1$ and $BOX_2$ (Figure \ref{fig4}, panels a-c). Moreover, passing  from $BOX_2$ to $BOX_3$ the PDFs of $D_p$ and $D_l$ 
become more similar, suggesting that proton inertia become less important in the exhaust of reconnection. 
The evolution of the conditioned average is interesting as well (Figure \ref{fig4}, panels b-d-f). Moving from $BOX_1$ to $BOX_3$ 
the conditioned average of $D_p$ increases for bigger values of the current density threshold.
This is more evident in the direct comparison shown in Figure~\ref{fig5}, where the 
normalized conditioned average grows from $BOX_1$ to $BOX_3$ (panel a). Similarly, the structures filling factor 
shows higher tails passing from $BOX_1$ to $BOX_2$ and $BOX_3$.

\section{Conclusions} \label{sec:2}

We have used a 3D kinetic numerical simulation to study the properties of the 
turbulence that develops in the outflow of magnetic reconnection using parameters 
typical of the Earth magnetotail. 
Simulations have shown that such turbulence is anisotropic, with 
large scales dominated by fluctuations whose wavevector is 
directed in the direction of the reconnecting magnetic field. 
Magnetic and electric turbulent energy spectra follows
two different power laws at scales smaller than the proton 
inertial length, with slopes which are in agreement 
with observations. 
Like the turbulent activity, the energy exchange between fields and particles is concentrated
in the outflows, where the strongest values of dissipation 
are found at the interface between the plasma ejected by reconnection 
and the ambient plasma.
Statistical analysis of the dissipation proxy confirms that the energy 
exchange between fields and particles occurs in small volume 
filling structure where the value of the current is much 
stronger than its root mean square. 
The current sheets produced by the turbulent activity compared
to their root mean square values are stronger in numerical 
simulation compared to observations. 
This difference could be due to the curlometer 
technique used to estimate current density from 
magnetic field measurements.
Moreover, we showed that in those places where the value of the current densities are
very high and which are not resolved by observations, the energy can also 
be transferred from particles to field. 
Finally, we showed that the properties of the turbulence
produced in the outflows varies in space
becoming more intermittent moving far from 
the X-point. 
We believe that these results could be used to better 
explain the upcoming MMS Mission observations
of the magnetotail.

\acknowledgments
The present work was supported by the NASA MMS Grants No. NNX08AO84G and NNX14AC39G, 
by the NASA Heliophysics Grand Challenge grant NNX14AI63G, 
by the Onderzoekfonds KU Leuven (Research Fund KU Leuven), 
by the Interuniversity Attraction Poles Programme of the Belgian Science Policy Office 
(IAP P7/08 CHARM) and by the DEEP-ER project of the European Commission. 
The simulations were conducted on NASA (NAS and NCCS) supercomputers, 
on the VSC Flemish supercomputing centre and on the facilities provided by PRACE research 
infrastructure Tier-0 grants. This research used resources of the National Energy Research 
Scientific Computing Center, a DOE Office of the Science User Facility supported by the 
Office of Science of the U.S. Department of Energy under Contract No. DE-AC02-05CH11231.

%\citep{huang2010scaling}

%\appendix
%\section{Appendix information}

%\bibliography{biblio.bib}
%\bibliographystyle{aasjournal}

\end{document}